# Designing Vibes in a **Science Museum:**

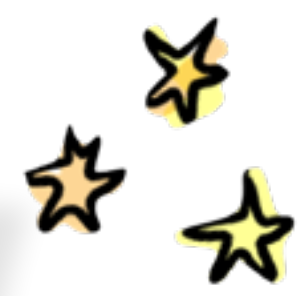

## from @Science to @🤗

**Derya Akbaba**
KTH Royal Institute of Technology,
Stockholm, Sweden
akbaba@kth.se

**Daniela Paz Moyano Dávila**
Linköping University,
Norrköping, Sweden
daniela.moyano-davila@liu.se

**Måns Gezelius**
Linköping University,
Norrköping, Sweden
mans.gezelius@liu.se

**Yin He,**
Linköping University,
Norrköping, Sweden
yin.he@liu.se

**Miriah Meyer**
Linköping University,
Norrköping, Sweden
miriah.meyer@liu.se

## Abstract

While feminist and critical data theories have long critiqued the use of data to uphold a positivist-informed view about science, few examples offer alternative methods to display scientific constructs. In response, we present *Data & Me*: an exhibit informed by feminist and critical data theories, which we designed and launched at a local science museum. *Data & Me* introduces museum visitors to data using a @🤗 vibe – a vibe that signals that data can be #slow, #handmade, and #personal. We designed this vibe to be noticeably different than the @Science vibe in the rest of the museum. Throughout our design process, we adapted visualization vibes as an analytic and generative tool in the context of a science museum. We present four design choices that enable the design of a vibe: visual, topical, material, and crediting. We discuss how our exhibit aligns with ongoing discussions about alternative research outcomes and calls for plurality in HCI

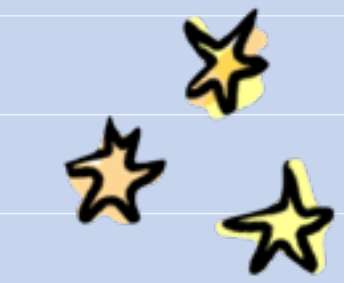

# Introduction

Science museums across the world serve as important spaces for educating and engaging the public in science, technology, engineering, and math. These science museums[1] – also referred to as science centers – have evolved from places that exhibit the instruments and objects of scientific study into experiences that make scientific concepts embodied, intuitive, and socially relevant. The interactive nature of science museums makes them interesting sites for HCI and visualization research[2], particularly for the design of interactive exhibits, learning systems, and virtual experiences.

We are a team of HCI researchers and designers situated in a research environment that includes a public science museum — the Visualiseringscenter C— that is collaboratively run by Linköping

Footnotes

1 Science museums have existed for hundreds of years as museums of curated scientific objects [7]. In the last century, however, innovative new approaches to science education gave rise to science centers with a focus on interactive experiences as well as contextualizing science within societal issues and challenges [39].

2 HCI researchers have extensively used museums as sites of research [44]. Much of the HCI research in museums touches on designing and evaluating new experiences with technologies, such as the use of VR [41], virtual spaces [29], and games [22]. More specific to science museums, visualization researchers have designed, deployed, and evaluated several notable interactive visualization exhibits: DeLVE [42] for experiencing the geological notion of deep time; LivingLiquid [34, 35] and Sea of Genes [15] for learning about the microscopic biology of oceans; and DeepTree [10, 26] for exploring the tree of life to better understand evolution.

University and the city of Norrköping, Sweden. The Visualiseringscenter C makes extensive use of data visualization technologies for science communication, and receives over 200,000 visitors annually. Many of the exhibits and experiences within the museum are created in collaboration with researchers from Linköping University, offering unique opportunities for both researchers and visitors to engage with public discourse of science. Within this context, we had the opportunity to design a new interactive exhibit that brings forward the question: What is data? [3]

Our exhibit is an alternative research outcome [47] that stemmed from a research project [2] that deployed data physicalization [4] workshops to local teens to engage them with data as a tool for personal reflection and visualizations as a form of self-expression [5]. We designed these workshops from a feminist epistemic stance, positioning data [6] as a produced artifact, entangled and inscribed with the context of its production. We wanted to similarly position data from this feminist perspective within our exhibit, and to communicate to visitors that data isn't just something that comes from fancy scientific equipment, but is something that anyone can make and use.

But how to communicate these specific socio-cultural messages within a science museum context? Recent work introducing *visualization vibes* [7] provides a conceptual framework for unpacking the ways that visualizations communicate the contexts in which they are expected to be used and the characteristics of the people expected to use them. In our design process, we explored ways to bring a feminist vibe into the Visualiseringscenter C by first analyzing the existing vibe of exhibits in the museum. The @Science vibe that we observed reflects traditional approaches [8] to communicating scientific topics, such as the use of official datasets and precise displays. We then worked to create a contrasting @🤗 vibe to signal an alternative perspective of science. Reflecting on our design process revealed a set of purposeful design decisions and descriptors that we employed in the new exhibit to implicitly communicate that everyone has the agency to produce and use data.

In this pictorial, we present a set of contributions for bringing feminist and critical perspectives into science museums. First, we conceptualize the @🤗 vibe that employs feminist and critical perspectives of data for explanatory visualizations through a set of design descriptors – #handmade, #playful, #personal, #relatable, #beautifullymessy, #collaborative, #alive, #care-full – and juxtapose it with the @Science vibe typical of science museums – #precise, #impressive, #official, #scientific, #technological, #curated. We actualize the @🤗 vibe in our second contribution, a new interactive exhibit called *Data & Me* that introduces the concept of data to a general audience. And third, we articulate four design choices that operationalize visualization vibes for considering intended social meanings when designing interactive exhibits in science museums: visual, topical, material, and crediting choices. This work illustrates the productive fluidity of *alternative research outcomes* [9] for offering opportunities to develop new research insights. We offer this work as part of an ongoing conversation in HCI and visualization to embrace plurality in research, design, and outreach.

3

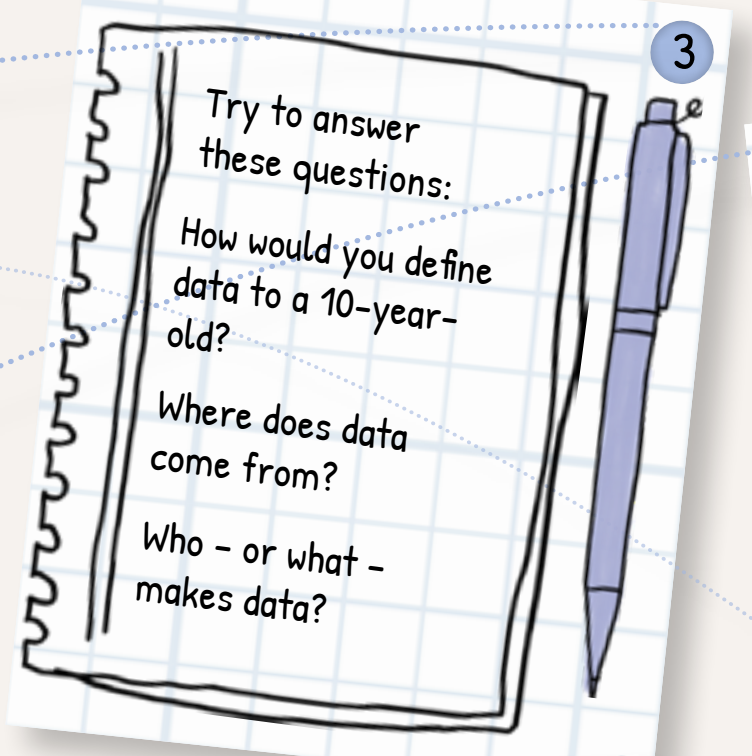

4

Data physicalization – the use of physical rather than digital materials to encode data [27] – has been used in workshops to engage a wide range of audiences on topics ranging from education [3, 28, 43] to creativity facilitation [13, 18] to community activism [9, 40].

5

Work by Lupi and Posavec, two designers who use hand-drawn data visualizations to communicate the mundane aspects of their lives to one another [33], explicitly rejects the idea of data as a purely precise and quantified endeavour. This rejection is further explicated in Lupi's data humanism manifesto [32].

6

Critical data studies and feminist scholars [12, 17, 21, 30] enumerate the many ways that data are designed artifacts that reflect hegemonic systems of power and privilege. Recent work relates these ideas directly to HCI [20] and data visualization [1] through a translation of entanglement theories.

7

The vibe of a visualization – or it's socio-indexical function – works to imply social and cultural origins and intentions of a visualization through elements like typography, color, chart type, and data complexity [36]. In a pair of studies, researchers documented that different visualizations have distinct vibes and that those vibes caused readers to make inferences about visualizations' social provenance [19, 36]. Furthermore, the vibes influenced the readers' reception to, and engagement with the visualizations based on their own social contexts.

8

Feminist scholars have, for some time now, called attention to the work that visualization conventions do such as those often used in communicating science. These critiques highlight how certain conventions – clean layouts, geometric designs, and two-dimensional viewpoints – reinforce that data are objective [24, 25] and a god-like view from nowhere [23].

9

Alternative research outcomes are activities and artifacts that emerge from HCI projects as researchers work to translate, communicate, or disseminate research findings beyond traditional academic audiences [47].

# Data & Me

The exhibit *Data & Me* consists of an interactive touch table that introduces visitors, over several digital screens, to data, how to data-ify objects in the world, and then how to construct visual representations of data. Descriptions on the touch table are complemented with interactive mini-games that illustrate the concepts being discussed. The digital touch table is paired with a large projected map of Norrköping, where visitors can place a small circular sticker on an overlaying acrylic board in response to questions about their experiences in the city, such as their favorite location or where they like to do their favorite activity.

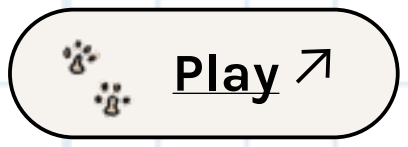

**What is Data** While visitors can interact with the exhibit in any way they like, the loose narrative structure on the touch table begins with a simple introduction to data as something personal and easily collected with simple tools like a pen and paper. The majority of screen space is given to a voting minigame, which demonstrates how their vote becomes a data point in a communal dataset.

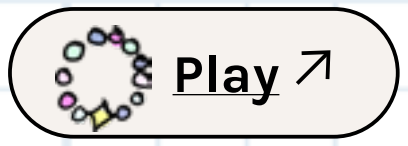

**Data Jewelry** The next screen outlines how real-world objects and experiences can be abstracted into data and visually crafted into representations like beads on a necklace. A tutorial exemplifies this point by walking through how food that we eat can be tokenized using beads.

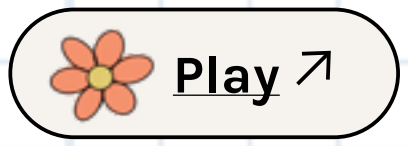

**Data Flower** On this screen, visitors can build their own personal visualization by playing a mini-game where they can choose emojis that represent activities they did during the week. Each click on an emoji representing an activity results in a new petal on a flower.

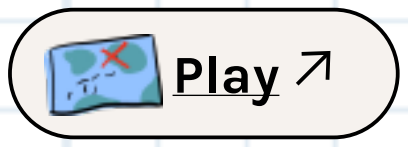

**Data Mapping** The final interactive screen invites visitors to respond to a question about their personal experiences in Norrköping by adding a sticker to the adjacent physical map.

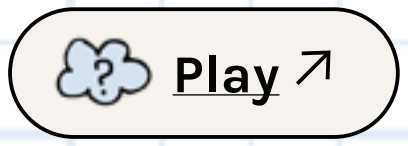

**About** We wrap up with the backstory of the exhibit and make visible the people and organizations involved with its creation.

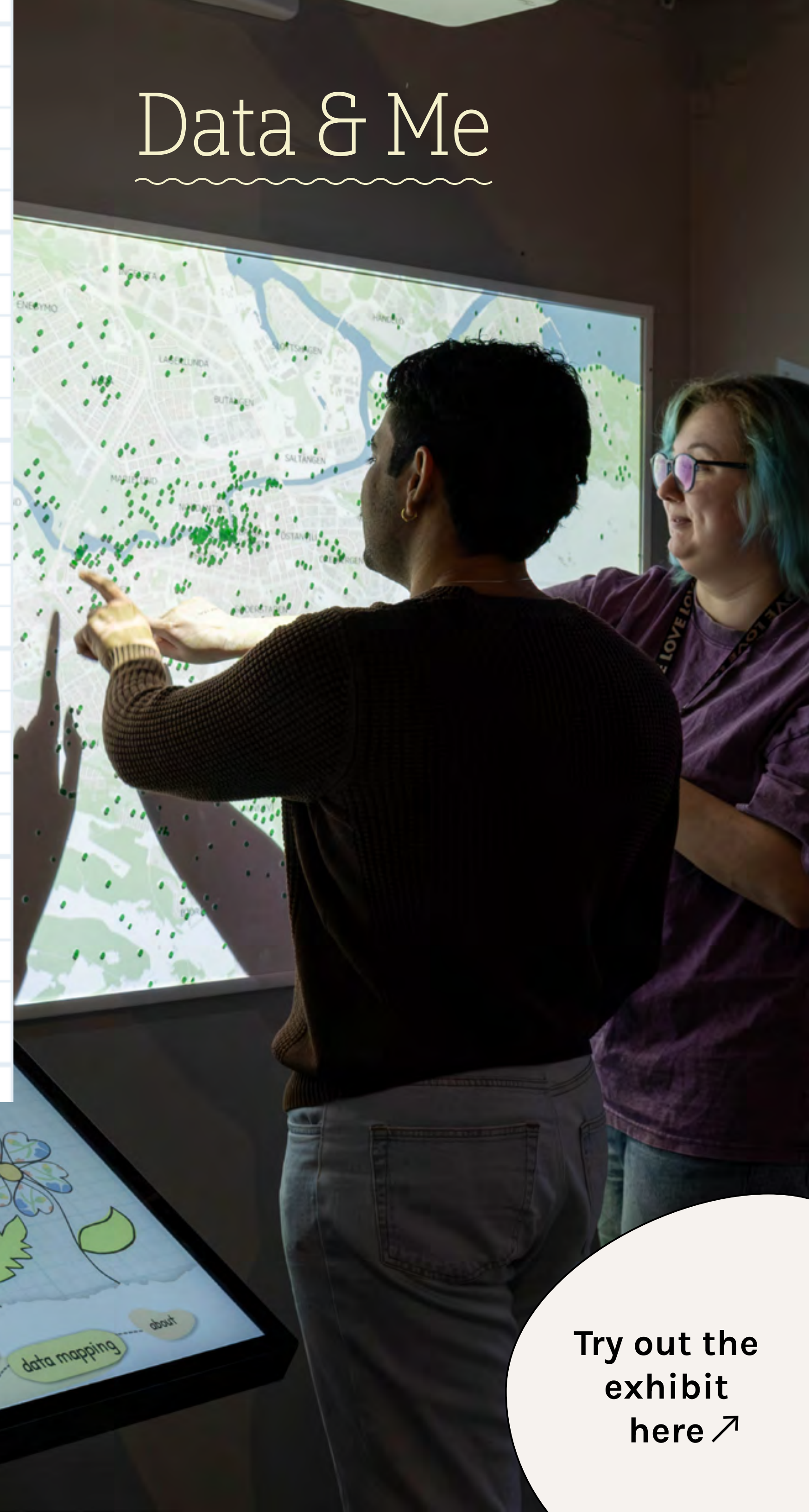

Try out the exhibit here ↗

# Vibe Check

In the following section, we articulate different choices we made to create a feminist @🤗 vibe for *Data & Me* that signals data as produced, entangled, and for everyone. We present four vignettes to illustrate how we translated feminist epistemology into visual, topical, material, and crediting choices. In each vignette, we highlight one of these choices and discuss the context for our decision-making. We juxtapose our @🤗 vibe choices with @Science vibe examples from the rest of the Visualiseringscenter C to make these decisions and their consequential vibes **pop!!**

To facilitate legibility, we name vibes and their descriptors using common social media conventions. Vibes are referenced using the @-sign and tagged using #'s. Our goal is to reinforce that vibes are situated and relational [36] — we named the vibe, but it could be named differently by others — and that descriptors are multiplicitous — there is more than one way to characterize a vibe. We summarize the @🤗 and @Science vibes in the diagram to the right, along with the descriptors we employed through our different design choices, presented in the following vignettes.

In keeping with vibes, in the following pages, we present the 4 vignettes with 2 pages each, designed intentionally to juxtapose one another. The first page of each vignette is stylized like the exhibit, representing @🤗. The second page of the vignette represents observations for the @Science vibe. We styled these @Science pages differently, using black and a more geometric typeface to mimic the rest of the science museum. That way, you too can read, compare, juxtapose, and get a feeling for the different vibes.

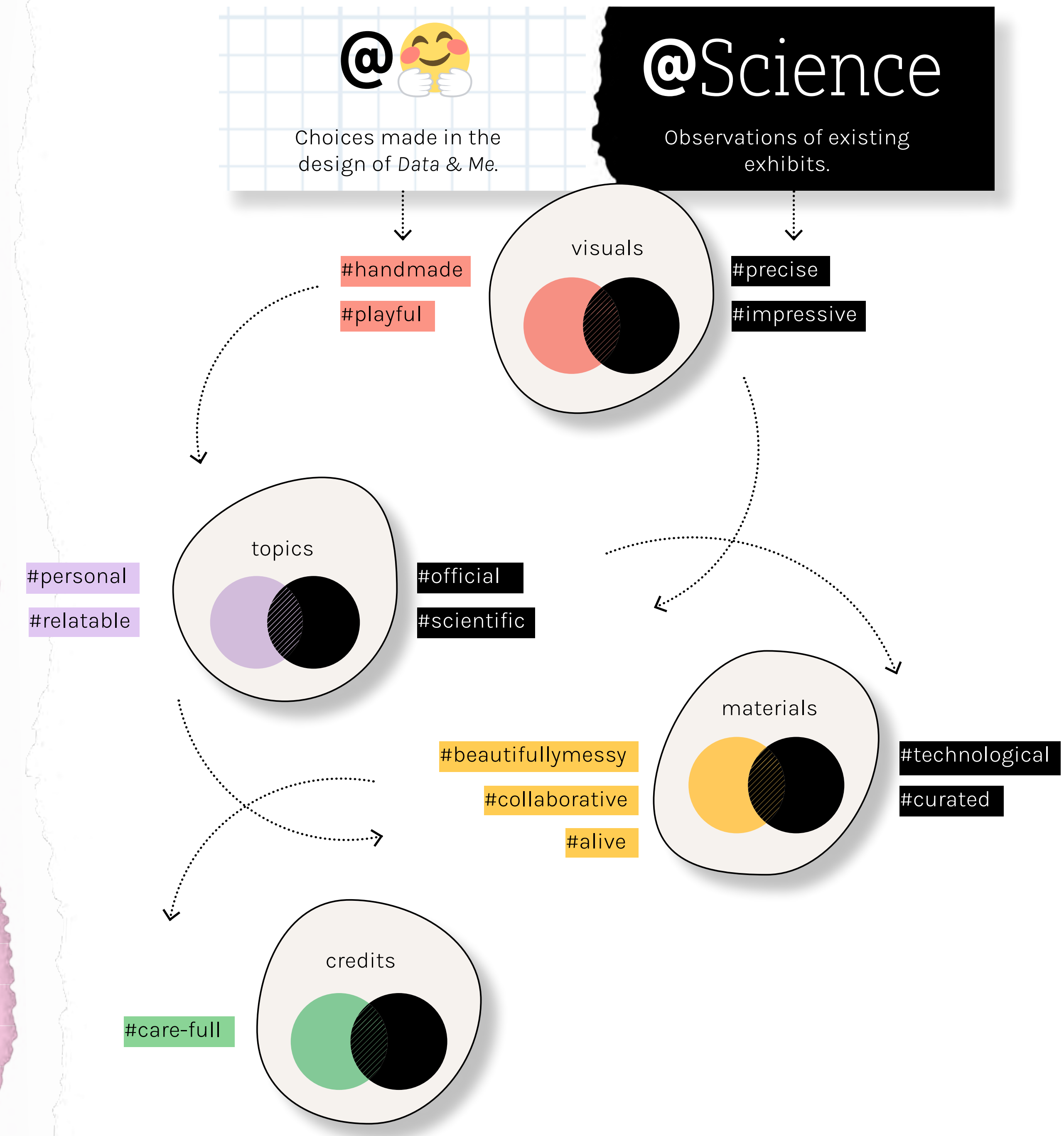

# Visuals

@😊: #handmade #playful

**Choices for *Data & Me***

We wanted viewers of the exhibit to feel like they could make and use data in their daily lives. Our process of designing the exhibit began with mood boarding, and we experimented with how to visually signal this vibe with different colors, geometries, fonts, and layouts. As we worked through these early iterations, we came to learn what we didn't want. We didn't want clean lines; it fit in too easily with critiques of visualization conventions that espouse neutrality [24].

At the time, the lead student on the project had been experimenting with #handmade illustrations in her presentations. She was doing this because the topic of her research did not lend itself to easy visuals, but she wanted some way to make the otherwise amorphous feminist theories approachable.

We tried out ideas to incorporate her #playful drawings into the exhibit with an overall visual style to match. The UX designer on our team warned us that it might be difficult to get approval for our illustrative visual design, because, as she put it: "it doesn't have a black background". What a vibe! Her comment drew our attention to the black background of the other exhibits in the science museum.

Still, our exhibit was approved, and we went to work creating hand drawn illustrations, along with a suite of #playful mini-games to craft our exhibit.

**Takeaway:** While there are many ways to visually represent a vibe, we chose a method that was authentic to the research team and an extension of our existing methods for communicating feminist research.

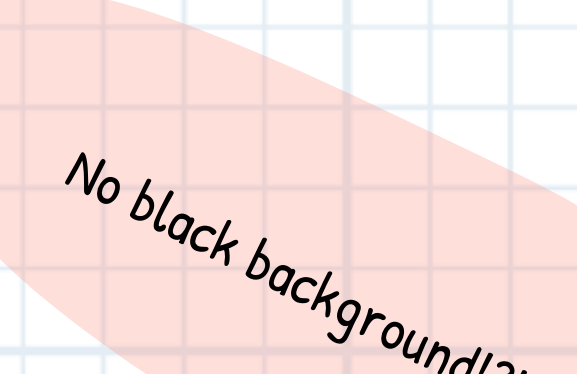

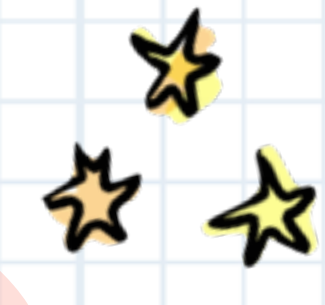

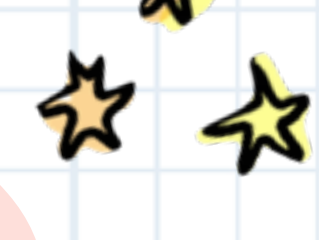
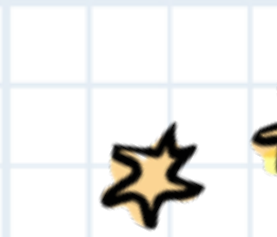
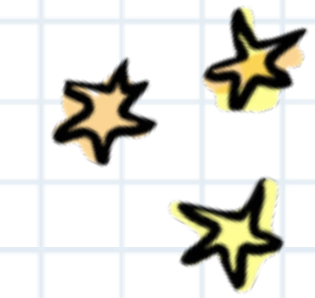
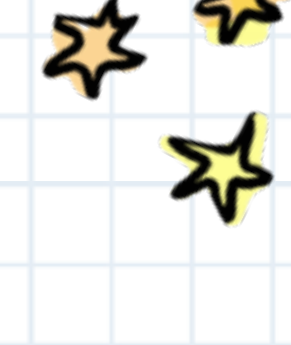

playful drawings

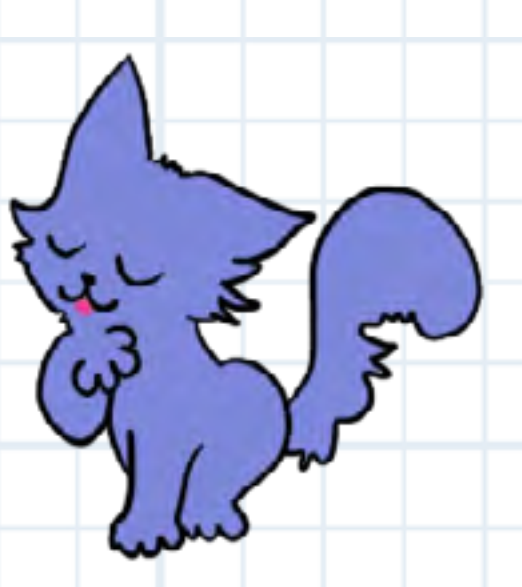

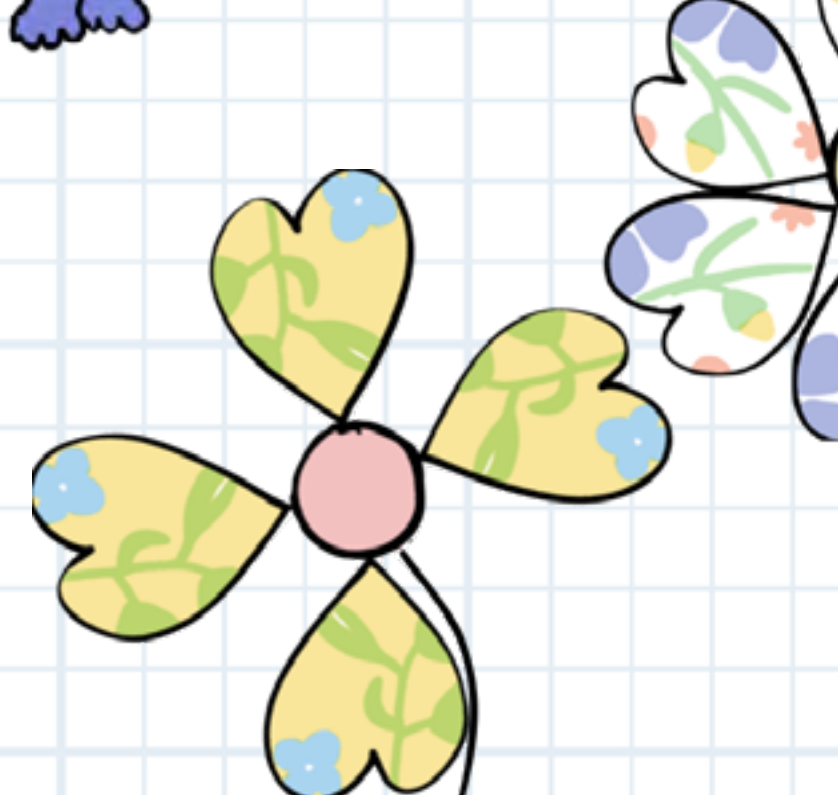

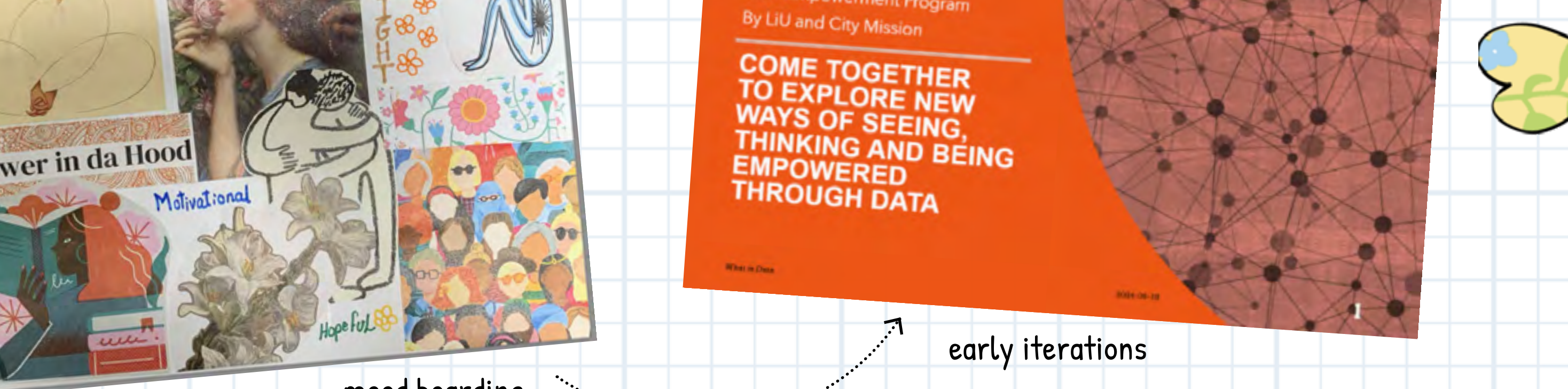

mood boarding

early iterations

@Science: #precise #impressive

# Visuals

**Observations of *OpenSpace***

The Visualiseringscenter C includes a dome theater for immersive experiences and films, the most unique of which is a tour of the universe using the OpenSpace software [11]. OpenSpace is a NASA-funded astrovisualization tool that renders #precise, interactive views of the entire known universe using data from observations, simulations, and space mission planning and operations.

Astronomical objects shown in OpenSpace are visualized using images taken by telescopes and satellites, 3D digital models of scientific instruments like the International Space Station, and planetary surfaces reconstructed from high-resolution images. The use of massive quantities of high-resolution data results in #impressive and realistic representations of space. These visuals reinforce science as a technology and data-driven field.

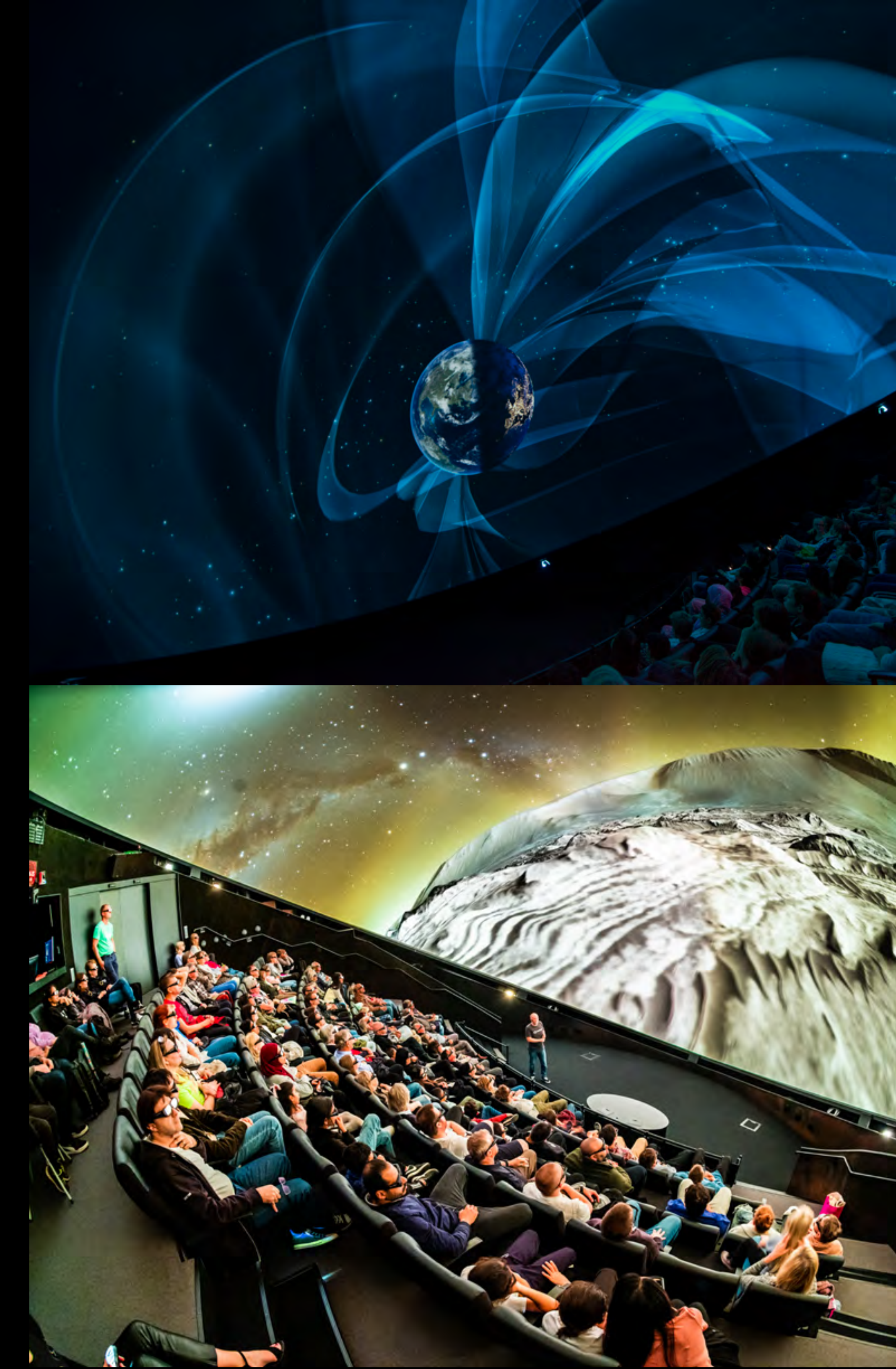

# Topics

@🤗: #personal #relatable

### Choices for *Data & Me*

Inspired by feminist critiques of science and the personal emphasis of data humanism [32], it was important to us to make the idea of data relevant, accessible, and interesting to visitors [38]. We wanted to show that data does not just come from fancy scientific instruments and equipment, but also from our daily lives. So for the exhibit, we chose data on relatable topics [33] that would resonate with our local community.

And everyone — well, perhaps not dentists — likes candy. This is especially true in Sweden where there is a tradition of buying candy from bulk containers on Saturdays. Candy is an important touchstone in modern Sweden society, and it is not just for kids, but for people of all ages. We include data about candy to bring forward the #personal potential of data.

Other deliberate topical choices within the exhibit include a voting activity to choose your favorite pet (cat v. dog), a visualization activity to make a flower based on what activities you did during the week, and a data physicalization activity to record your experiences in the city using a sticker. All of these examples are intentionally #relatable.

**Takeaway:** We interpreted the personal relevance of data as an important value for feminist visualization, and used it as a guiding theme when choosing examples for the exhibit.

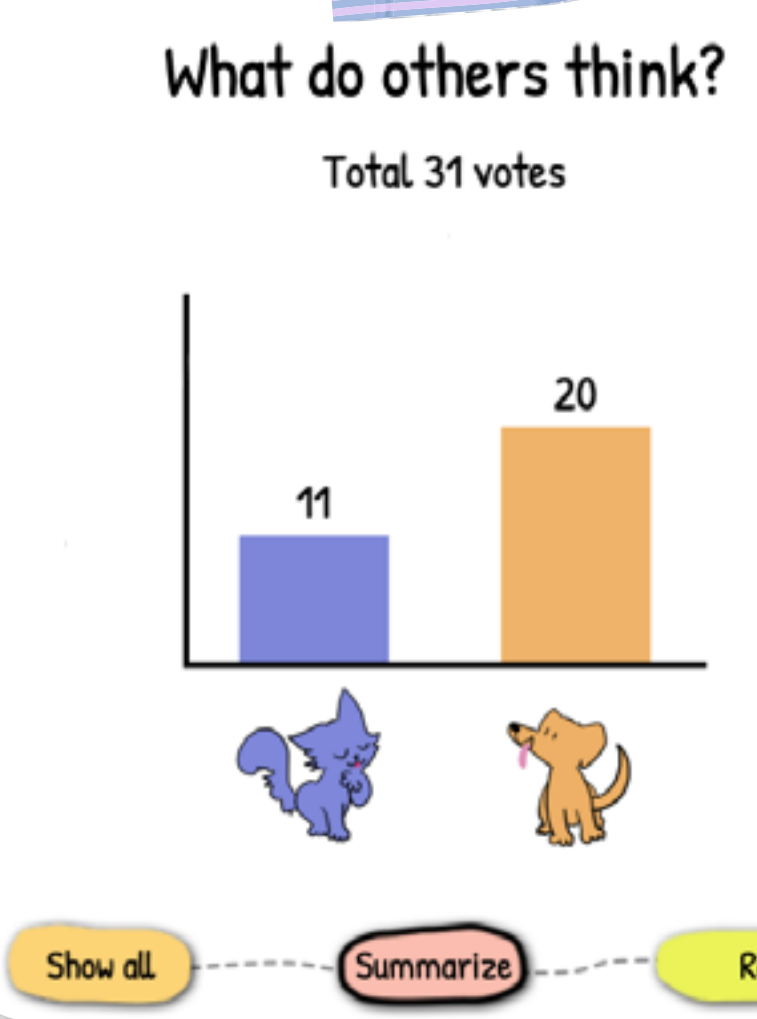

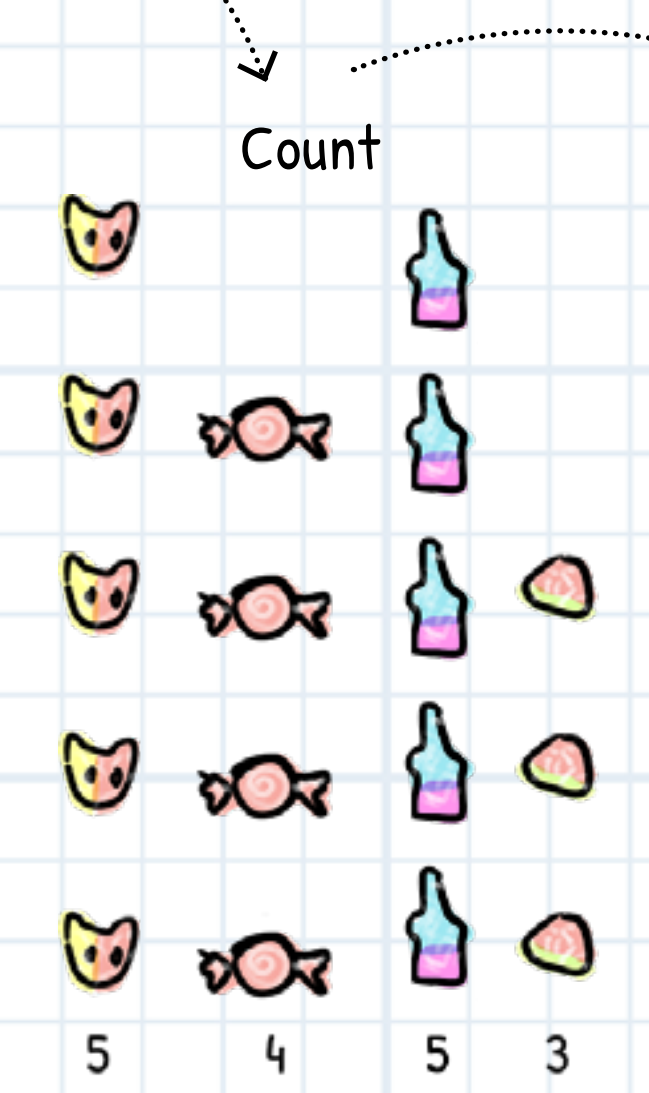

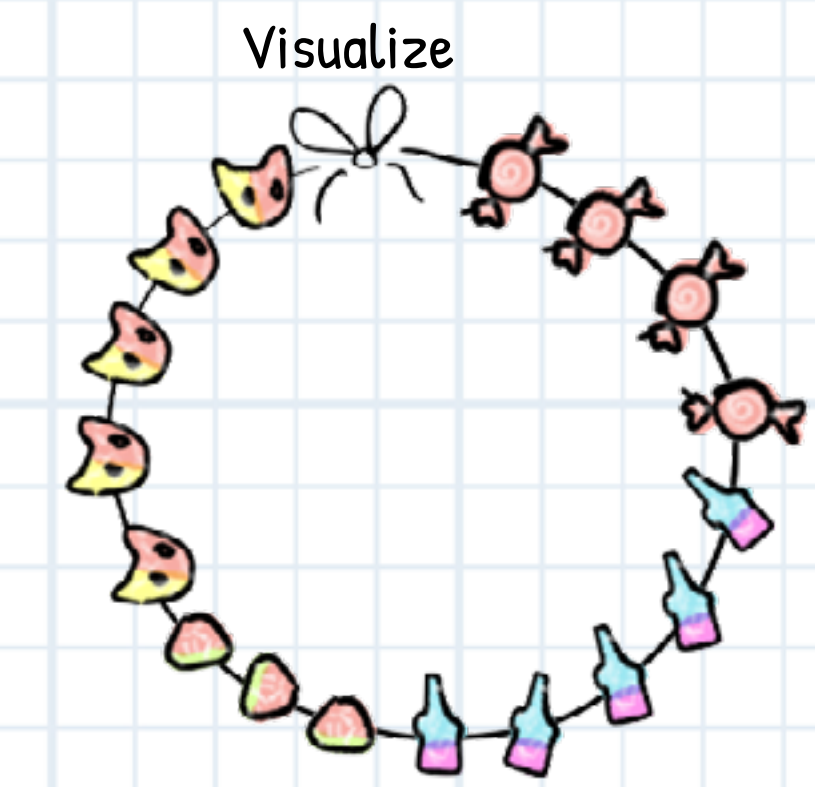

@Science: #official # scientific

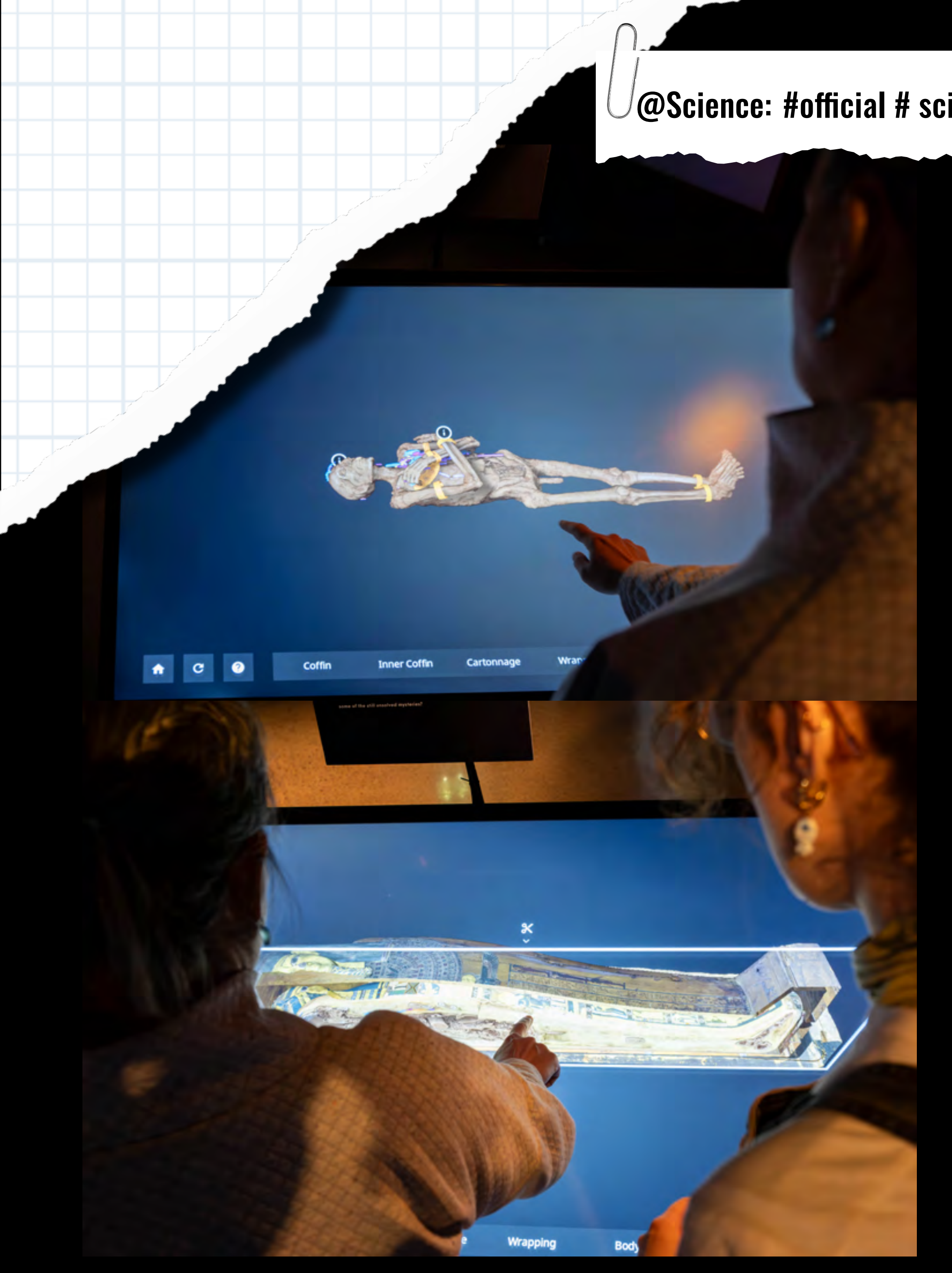

# Topics

**Observations of *The Virtual Autopsy***

A guiding principle in the Visualiseringscenter C is exploranation [45], the idea that the tools of science can also be the tools of scientific communication. Bringing #official, #scientific datasets together with advanced visualization and interaction technologies provides audiences opportunities to explore data in the same ways scientists do.

For example, in a range of scientific investigations — like virtual autopsies and archaeological studies [31, 46] — scientists visualize datasets generated from CT scanners on large-scale multitouch tables. At the Visualiseringscenter C, visitors use a simplified version of these interfaces to explore datasets like the 3D model of the Neswaiu mummy. This dataset was generated from scans of the mummy and the surrounding sarcophagi, allowing researchers to study the remains without disturbing burial wrappings. Using interactions on the touchtable — like slicing through the volume and rendering different materials — visitors too can explore the Neswaiu dataset, including views of amulets encased within the wrappings.

# Materials

@: #beautifullymessy #collaborative #alive

**Choices for *Data & Me***

We were interested in bringing forward data as dynamic in its making [4], materially connected to the physical world [37], and open to all [8, 16]. To do this, we introduced a data physicalization map that invites museum visitors to place a sticker on their favorite location in Norrköping. As more people contribute their stickers to the map and the favorite spots become more visually salient, traces of visitors' experiences within the city come into view. As does data graffiti, removed stickers, and other evidence of people interacting with the map. The result is a #beautifullymessy visualization that reflects the slow process of #collaborative data collection.

Given the reliance on digital technology in the center and potential for things to go wrong with physical materials, there was initial pushback to our idea of using stickers rather than digital marks. But we wanted to challenge the idea that data only exists as bits and bytes, and to question what gets to count as data, and who gets to make it. We wanted data to be #alive.

**Takeaway:** Considering the material aspects of the exhibit in relation to the rest of the science museum, we intentionally chose an underrepresented modality for representing data.

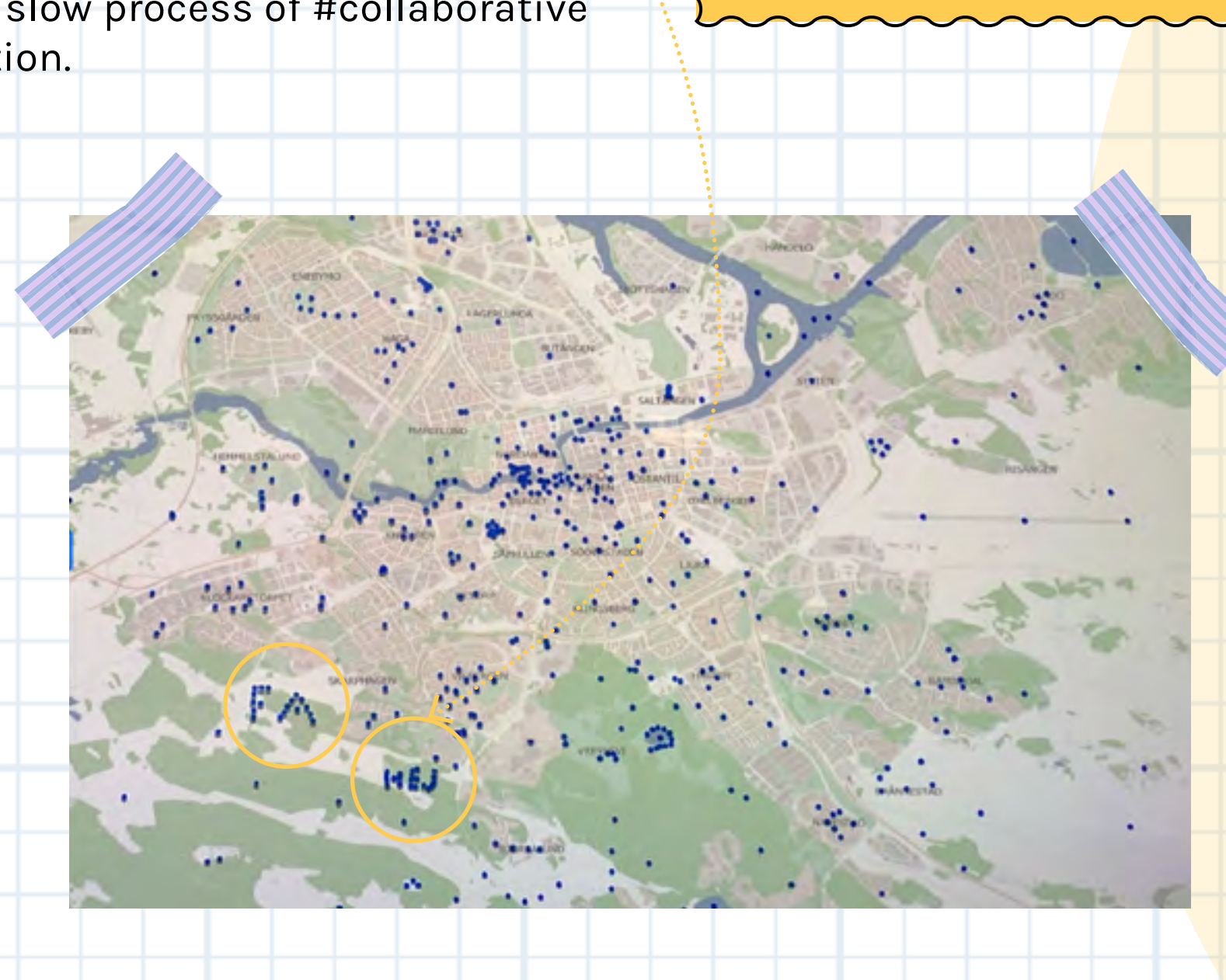

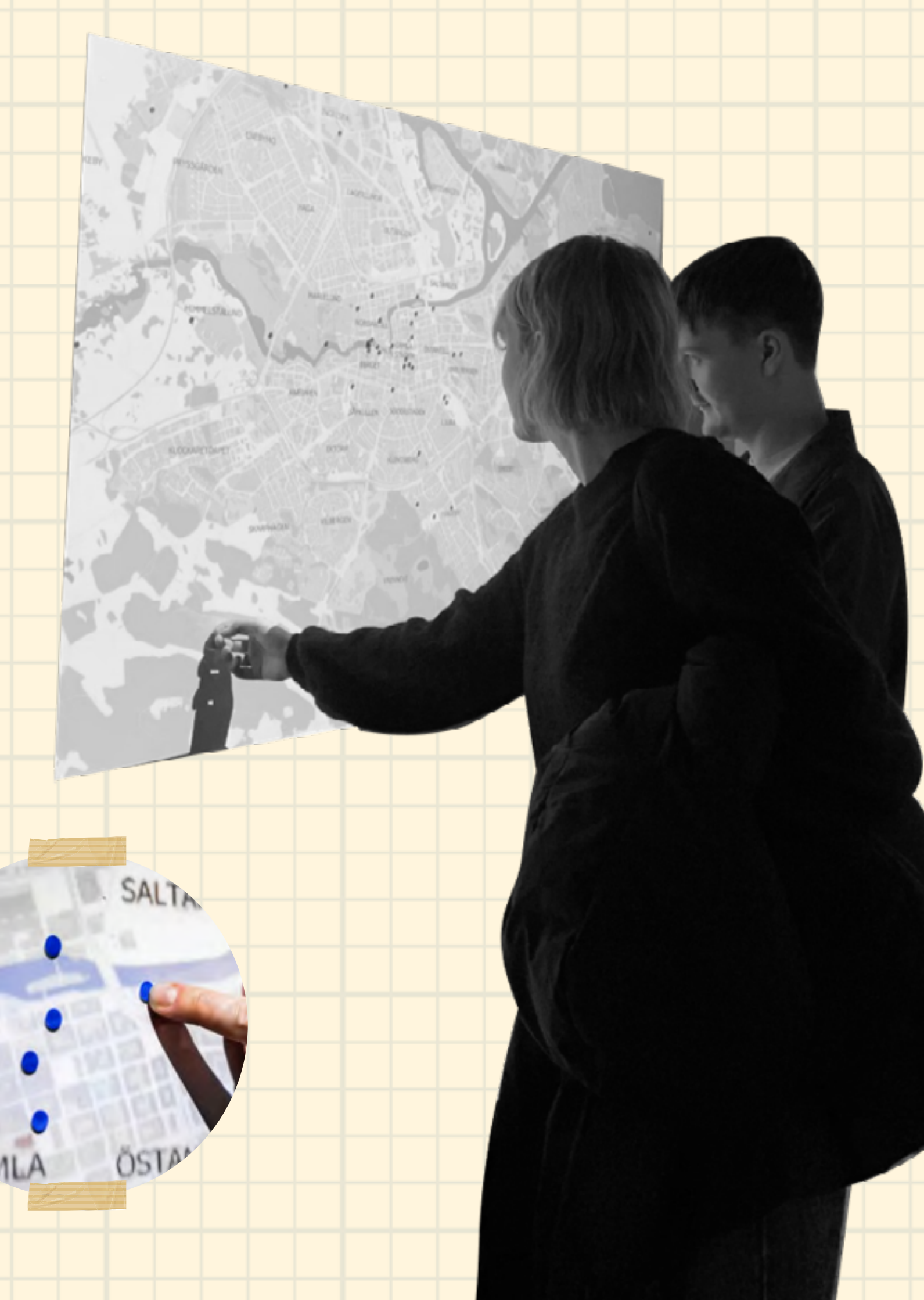

@Science: # technological #curated

# Materials

**Observations of *Visual City***

The Visual City exhibit uses a 3D printed map of Norrköping as a projection surface for visualizing a range of datasets about the cityscape, including historic boundaries, the potential impacts of climate change, and the results of urban infrastructure like noise levels and transportation routes. These datasets come from satellite imagery, meteorological simulations, and urban plans as part of a larger #technological project to create a digital twin of Norrköping. The 3D-map was generated from fly-over videos taken by drones and processed using custom, in-house software to computationally construct and print a high-resolution model of the urban landscape. The technology underlying the exhibit is a proof-of-concept around the possibilities in automating the creation of digital twins, a process that otherwise requires labor-intensive work of manually mapping out land use. The Visual City exhibit provides Norrköping's urban planners and museum visitors with a #curated, bird's-eye-view of the city's past, present, and possible futures.

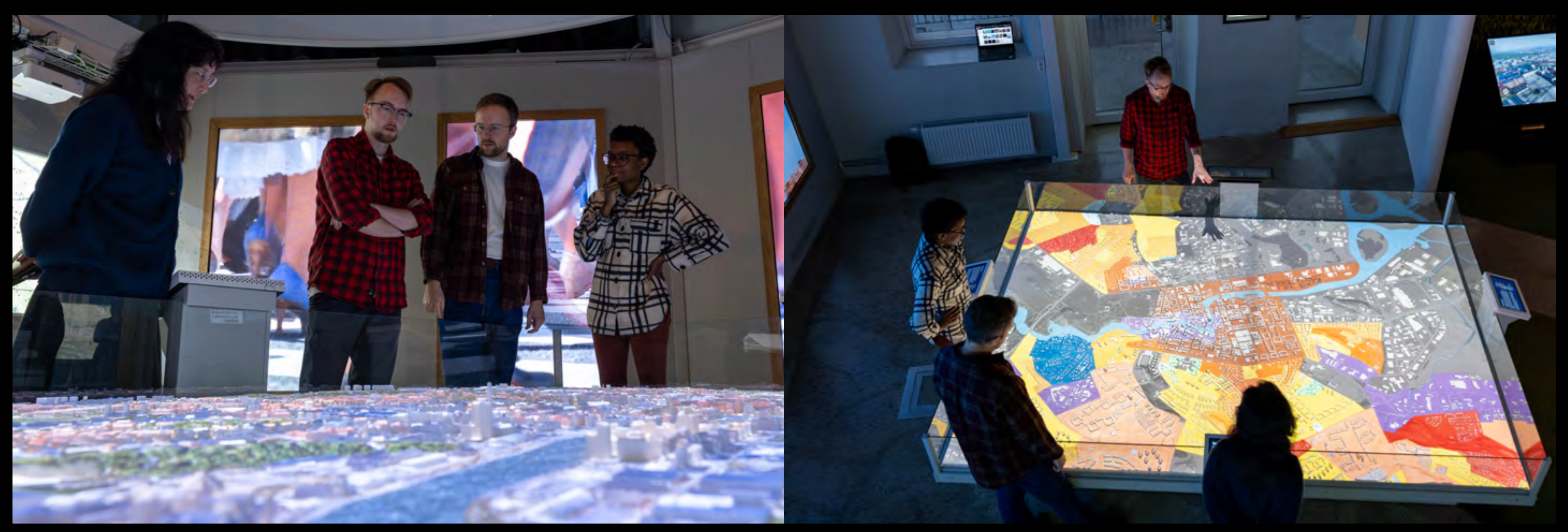

# Credits

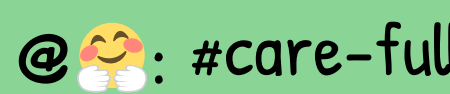

Derya Akbaba
Contribution

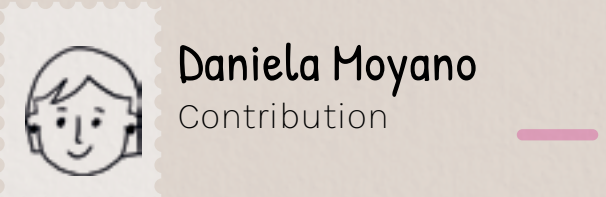
Daniela Moyano
Contribution

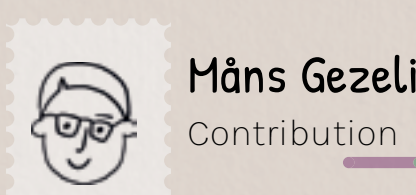
Måns Gezelius
Contribution

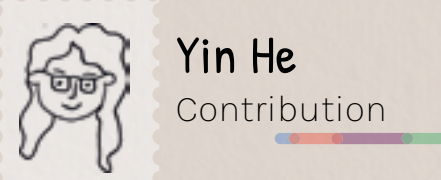
Yin He
Contribution

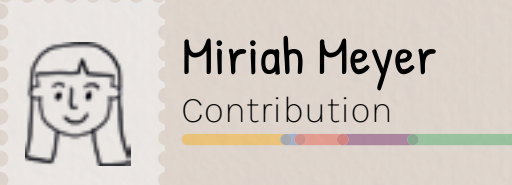
Miriah Meyer
Contribution

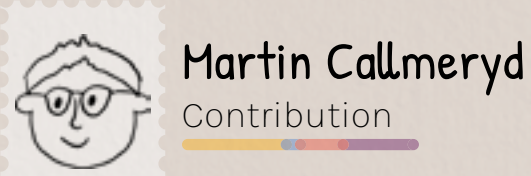
Martin Callmeryd
Contribution

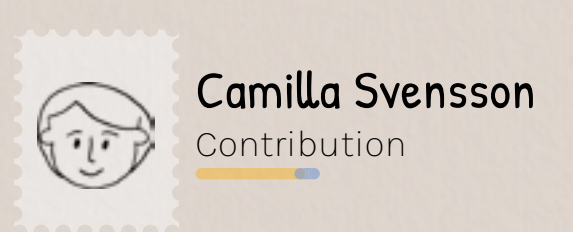
Camilla Svensson
Contribution

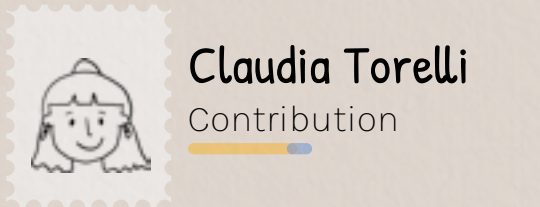
Claudia Torelli
Contribution

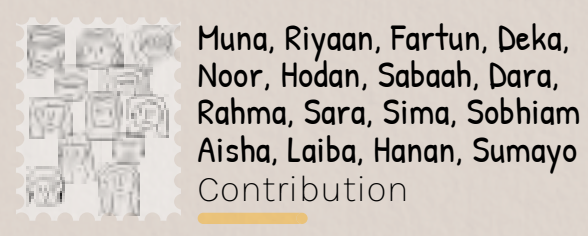
Muna, Riyaan, Fartun, Deka, Noor, Hodan, Sabaah, Dara, Rahma, Sara, Sima, Sobhiam Aisha, Laiba, Hanan, Sumayo
Contribution

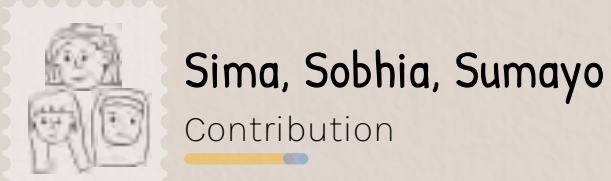
Sima, Sobhia, Sumayo
Contribution

### Choices for *Data & Me*

The exhibit did not begin as a dedicated research project, but rather was conceived as an alternative way to make visible the work of participants and collaborators from an earlier research study [2]. This previous study was done in collaboration with managers at a local nonprofit and brought data physicalization workshops to teen participants. And so, bringing the collaboration to the Visualiseringscenter C was an important way to not only bring visibility to the participants, but also to support the organization in demonstrating impact for future funding.

The initial ideas for the exhibit drew directly from activities in the workshops and mood-boarding from several of the participating teens. Further iterations on the exhibit with museum staff, however, gradually moved it away from just the story of the research collaboration and instead towards a more general exhibit about data. As we were wrapping up the design of the exhibit and thinking about how to acknowledge the labor [17] of the research team, nonprofit organization, local teens, and museum staff, it was unclear how to do this in an interactive exhibit.

We looked in the Visualiseringscenter C to find out how other exhibits handle crediting, and found out that none of them do. This insight has since spurred a companion research project exploring norms of invisibility across science museums for interactive exhibits. More locally, it inspired us to include a #care-full about-page and name the people behind the ideas in the exhibit.

**Takeaway:** There will be no page representing credits for @Science. Finding things that are missing, in this case, credits, is one way of creating a different vibe.

### Phases of the exhibit

- Nov 2023 – May 2024: Inspiration for exhibit began with a research project working with managers at a local nonprofit and teen participants.
- May 2024 – July 2024: Brainstorming ideas to bring visibility to nonprofit and teens.
- July 2024 – Oct 2024: First design iteration with manager of nonprofit, museum staff, and research team to generalize ideas from joint workshops.
- Oct 2024 – March 2025: Second design iteration with museum research developer to include interactive mini-games.
- March 2025 onward: **Public launch of Data & Me** — Life of the exhibit
- Nov 2025: Pictorial Writing

Nov 2023 · May 2024 · June 2024 · July 2024 · Sept 2024 · Oct 2024 · Feb 2025 · March 2025 · Nov 2025

# Discussion

## From ARO to Research Insight

Alternative research outcomes (AROs) are a recently proposed concept for translating and disseminating research beyond academia [47]. We see this conceptualization as an invitation to researchers to experiment in new ways of communicating findings and interacting with diverse audiences. We present *Data & Me* as an ARO, originally conceived as an exhibit to highlight our work with a local nonprofit and teens and make their contributions visible. As a research team, this was our first experiment in creating an exhibit for the public in a science museum.

The *fluidity* of the exhibit context – the flexibility and open-endedness of AROs – allowed us to freely pursue a vision for our treatment of data for the public. Our desire for an inclusive and approachable portrayal of data, as expressed in the **@**🤗 vibe, is rooted in our readings of feminist critiques and theories of science. Embodying those perspectives, we were able to design a new positionality towards data and generate an alternative view for the public. The ARO gave us the space to experiment without worrying about precise translations and theoretically correct arguments, and to create something new.

And yet, despite the explicit intention of AROs to disseminate beyond academia, here we are writing a pictorial about our work. We found that discussions with colleagues about *Data & Me*, and specifically about our intentions to design a feminist exhibit about data, prompted interest and questions about what we did. These questions pushed us to reflect more deeply about our process and what we learned, and ultimately led our articulation of the contributions we present here. This leaves us with a tension: In what ways can AROs contribute to our body of academic knowledge? And, should they?

## Juxtaposition as Feminist Praxis

Since the introduction of feminist theories to HCI, there have been numerous calls to move beyond critique and toward a feminist practice [5, 6, 14]. And yet, we struggle in our own research to move beyond the resonant critiques from feminist theory and to see a generative alternative. We felt this tension early on in our design process, where our admiration for the Visualiseringscenter C clashed with scholarship from critical and feminist data studies. We knew we wanted to contribute a different vision for how science could look, but we also loved the vision that already existed. In this way, critique stopped short for us.

While writing this paper, we held many discussions about whether we were truly upholding plurality when representing our work and the work of the center. We focused on *juxtaposition* in this pictorial as a productive alternative to build a feminist objectivity [23]. We embraced the concept of visualization vibes [36] as it allows us to discuss the different socio-cultural messaging of exhibits across the Visualiseringscenter C, without necessitating a value judgment. In our exhibit and this pictorial, we demonstrate that both representations — **@**🤗 and **@** Science— belong in a science museum. We see this as a step toward representations of science that are multiplicitious and inviting to a broad range of audiences.

## Future Directions

While we have some preliminary engagement metrics for the exhibit, what we are most interested in is understanding the impacts of vibes on museum visitors' impressions of science. Perhaps **@**🤗 resonates more with young girls than boys, and sparks new interests in STEM? Maybe the content is more accessible to families, and leads to inter-generational discussions about data? We intend to evaluate and study the exhibit more formally in future work.

We are also interested in seeing how the choices presented in this pictorial manifest and transfer to other exhibits and museum contexts. While recognizing the situated and contextual nature of vibes, we are also eager to develop more formal methods for other researchers to operationalize feminist concepts in public outreach work.

## Acknowledgments

This work is funded in part by the Swedish Research Council (Grant No. 2024-05726); by the Wallenberg AI, Autonomous Systems and Software Program (WASP) funded by the Knut and Alice Wallenberg Foundation; and by the InspireLab at KTH. This work would not be possible without the participation of Östergötland Stadsmission, Power in da Hood, Martin Callmeryd, Camilla Svensson, and Claudia Torelli. Or without the support of management and staff at the Visualiseringscenter C. And of course, thank you to our supportive lab mates at Visualization and Interaction Design Lab and colleagues who took pictures and posed for photos, you all rock!

## Photo Credits

Pg. 3 Vsevolod Suschevskiy
Pg 6. Thor Balkhed
Pg. 8 Vsevolod Suschevskiy
Pg 9. Derya Akbaba
Pg 10. Vsevolod Suschevskiy
'paper page clip' Designed by vectorpouch / Freepik
'collection torn ripped paper' Designed by starline / Freepik
'coleccion de cintas realistas' Designed by pikisuperstar / Freepik